\documentclass[twocolumn,amsmath,amssymb,prl]{revtex4}

\usepackage{graphicx}
\usepackage{dcolumn}
\usepackage{bm}

\begin{document}

\preprint{APS/123-QED}

\title{Reply to Comment on ``Wigner rotations and an apparent paradox in relativistic quantum information''}

\author{Pablo L. Saldanha}\email{saldanha@fisica.ufmg.br}
\affiliation{Departamento de F\'isica, Universidade Federal de Minas Gerais, Caixa Postal 701, 30161-970, Belo Horizonte, MG, Brazil}
\author{Vlatko Vedral}
\affiliation{Department of Physics, University of Oxford, Clarendon Laboratory, Oxford, OX1 3PU, United Kingdom}
\affiliation{Centre for Quantum Technologies, National University of Singapore, Singapore}
\affiliation{Department of Physics, National University of Singapore, Singapore}


\date{\today}


\begin{abstract}
As Lanzagorta and Crowder have shown in their \textit{Comment} [Phys. Rev. A \textbf{96}, 026101 (2017)], the linear application of the Wigner rotations to the quantum state of two massive relativistic particles does not entail the instantaneous transmission of information as we concluded in our paper [Phys. Rev. A \textbf{87}, 042102 (2013)]. But in the new version of our paper in arXiv [arXiv:1303.4367v2] we show (and solve) another paradox that is generated by the linear application of the Wigner rotations to a system with a single relativistic particle. So the conclusion of our paper that we cannot in general linearly apply the Wigner rotations to the quantum state of a relativistic particle without considering the appropriate physical interpretation is still valid, although the paradox presented in that paper is inappropriate. 
\end{abstract}

\maketitle

In Ref. \cite{saldanha13} we have discussed that the linear application of the Wigner rotations to the state of a massive relativistic particle in a superposition of counterpropagating momentum states leads to a paradox that could entail the instantaneous transmission of information between two separated parties.  Lanzagorta and Crowder \cite{com} argued that we arrived at this conclusion due to a miscalculation in Eqs. (9) and (10) from our paper \cite{saldanha13}, that did not consider the phases included in Eq. (9) from Ref. \cite{com}. It is embarrassing to admit that Lanzagorta and Crowder are right. Upon correcting the calculations, the conclusion is that even if the linear application of the Wigner rotations is a valid procedure it would not be possible for Alice to transmit a bit of information to Bob in the way that we describe in our paper, as they show in Ref. \cite{com}. 

However, there are other paradoxes that arrive with the linear application of the Wigner rotations, as we show in the new version of our paper in arXiv \cite{new}. The linear application of the momentum-dependent Wigner rotations to the quantum state of a massive relativistic particle in a superposition of counter-propagating momentum states in combination with a general model for particle detection leads to the conclusion that the probability of finding the particle at different positions would depend on the reference frame \cite{new}. But the probability of finding the particle around some position cannot depend on the reference frame. A solution of the paradox is given based on the physical interpretation of the Wigner rotations given by us \cite{saldanha12a} and on a discussion about the preparation method of the quantum state of the particle. We conclude that the Wigner rotation depends on the preparation method, such that with a change of the reference frame the spin transformation of a state in a superposition of different momenta is not necessarily equivalent to the linear application of the momentum-dependent Wigner rotation to each momentum component of the state, a conclusion that solves the paradox \cite{new}.

So the conclusion of Ref. \cite{saldanha13} that we cannot in general linearly apply the Wigner rotations to a quantum state without considering the appropriate physical interpretation is still valid, although the paradox presented in this paper is inappropriate. 

P. L. S. acknowledges the Brazilian agencies CNPq, CAPES, and FAPEMIG for the financial support. V.V. acknowledges financial support from the Templeton Foundation, the National Research Foundation and Ministry of Education in Singapore and the support of Wolfson College Oxford.

\end{document}